\documentstyle[aps,pre]{revtex}

\def\eps{\epsilon}
\def\al{\alpha}

\def\la{\langle}
\def\ra{\rangle}
\def\cT {{\cal T}^{(p)}}
\newcommand {\be}  {\begin{equation}}
\newcommand {\bea} {\begin{eqnarray} \nonumber }
\newcommand {\ee}  {\end{equation}}
\newcommand {\eea} {\end{eqnarray}}
 \def\(({\left(}
 \def\)){\right)}

\draft

\begin{document}

\title{ The response of glassy systems to random perturbations: A bridge
between equilibrium and off-equilibrium}
\author{Silvio Franz\cite{Silvio}}
\address{Abdus Salam International Center for Theoretical Physics,\\
Strada Costiera 11, P.O. Box 586,
I-34100 Trieste (Italy)}
\author{Marc M\'ezard\cite{Marc}}
\address{Institute for Theoretical Physics,\\
 University of California Santa Barbara, CA 93106-4030 (USA)\\
and CNRS, Laboratoire de Physique Th\'eorique de l'ENS  (France)}
\author{Giorgio Parisi\cite{Giorgio}}
\address{Dipartimento di Fisica and Sezione INFN,\\
Universit\`a di Roma ``La Sapienza'',
Piazzale Aldo Moro 2,
I-00185 Rome (Italy)}
\author{Luca Peliti\cite{Luca}}
\address{Dipartimento di Scienze Fisiche and Unit\`a INFM,\\
Universit\`a ``Federico II", Mostra d'Oltremare, Pad.~19,
I-80125 Napoli (Italy)}
\date{\today}
\maketitle
\begin{abstract}
{%
We discuss the response of aging systems with short-range
interactions to a class of random perturbations. Although
these systems are out of equilibrium,
the limit value of the free energy at long times is equal
to the equilibrium free energy. By exploiting this fact,
we define a new order parameter function, and we
relate it to the ratio between response and fluctuation,
which is in principle measurable in an aging experiment.
For a class of systems possessing \textit{stochastic stability},
we show that this new order parameter function is intimately
related to the static order parameter function, describing the
distribution of overlaps between clustering states. The same
method is applied to investigate the geometrical organization
of pure states. We show that the ultrametric organization in
the dynamics implies static ultrametricity, and we relate
these properties to static \textit{separability}, i.e., the property
that the measure of the overlap between pure states is
essentially unique. Our results, especially relevant for
spin glasses, pave the way to an experimental determination
of the order parameter function.
}
\end{abstract}

\section{Introduction}
Ergodicity breaking in glassy systems is well understood within
mean-field theory \cite{BY,MPV,FiHe,APYoung}.  In this context the
description of thermal equilibrium has several peculiar aspects.
While in non-disordered systems it is usually possible to select pure
phases via boundary conditions or suitable external fields, this is
problematic in glassy systems, since one does not have in general the
a priori knowledge of the low free-energy configurations.
Consequently, the Gibbs measure for a large finite system will
generically turn out to represent a mixed state.  This phenomenon was
first discovered and understood in the mean-field theory of spin
glasses, and was named ``replica symmetry breaking'' (RSB) from the
technique originally used to analyze these models. However it is not
specific to disordered system, and recent theoretical studies have
shown that it appears in first-principle computations of the
vitrification of simple liquids \cite{vetri-rsb}.  Numerical
simulations unambiguously show that detailed predictions of the
mean-field scenario apply to a number of finite dimensional systems
\cite{romani-young}.  An experimental verification of the most
characteristic predictions of the theory in real materials is however
still missing.  The main difficulty lies in the fact that it appears
impossible to measure directly the basic quantities involved in the
theory.  One of the most important predictions of the theory is the
existence of infinitely many pure states, which depend stochastically
on the realization of the quenched disorder \cite{MPV}---if
present---and on system size \cite{MPV,ns}. In this situation one is
forced to adopt a probabilistic description of the Gibbs measure, and
to study the statistical properties of pure states.  A first
fundamental characterization is given by the ``overlap probability
function'' $P(q)$ \cite{par83}, which is the probability distribution
function of the codistance among equilibrium states, averaged over the
disorder and/or the system size.  However, its direct measurement is
opposed by two basic difficulties.  First, it would require a detailed
knowledge of the microscopic configurations of the system.  Second,
glassy systems are never at equilibrium on experimental time scales.
In this paper, we show that linear response theory allows to define a
new function $\tilde P(q)$ in terms of  a set of
responses to some  appropriate
perturbations in the Hamiltonian (i.e.,  generalized
susceptibilities), and to relate them in turn to some dynamical
functions which can be experimentally measured.
We then argue that,
for a large class of systems, $\tilde P(q)$ is directly related to the
equilibrium order parameter function $P(q)$.  These systems, which we
shall call ``stochastically stable'', are such that their
``metastate''
(i.e.,  the probability distribution of the pure states) is stable
under some random local perturbations of the Hamiltonian.

Linear response theory, i.e., the study of the effects of small
perturbations on physical systems, has
a long history in statistical physics.
One of its fundamental results is the
fluctuation-dissipation
theorem (FDT), which establishes, under the hypothesis of thermal
equilibrium, the equality of two
quantities of a priori very different nature: the response to a
small disturbance and the
correlation function describing the statistical fluctuation of the
disturbed variables.  The theorem
is of course of fundamental importance: in equilibrium conditions one
can get information about the
fluctuations of the system just by studying its response to a
perturbation.  This relation, however,
has no reason to hold in general  off-equilibrium situations, where
the response and the
correlation functions are unrelated.  In recent works, in fact,
several mean-field models of spin glasses and glasses have been studied,
in which the
asymptotic violation of time-translation invariance appears together
with the violation of the equilibrium relation between response
and fluctuations.  Slow,
history-dependent dynamics,
usually referred to as
\textit{aging}, is a constitutive property of glassy systems
\cite{struik}. It
has been observed in a number of experimental systems, which go from
polymeric glasses to charge-density waves, and in the
numerical simulations of practically all the
glassy systems where it has been
searched.  We shall mostly refer in this paper to spin glasses as the
prototypes of aging systems \cite{age-exp}
although our considerations can be applied to any glassy system.  We
notice in particular that one could discuss in a similar way
some
systems which are slightly out
of equilibrium due to the presence of a slowly time-dependent
Hamiltonian or small non-conservative forces\cite{CKLP,teff},
but we shall not develop
this discussion here.

In a typical aging experiment one monitors the relaxation of a
spin glass at low temperature, after a
quench from the high-temperature phase at an initial time.  One of the
most striking features of
aging dynamics is that even for the longest observation times the system
does not reach a stationary
state.  One can concentrate on two classes of observables: ``one-time
observables'', that depend on
the configuration of the system at a single time, and ``two-time
observables'', like correlation and
response functions, that depend on two times.  Due to the relaxational
nature of the dynamics, one-time observables
tend to time-independent limits.  The off-equilibrium
nature of the dynamics is
manifest in the behavior of some two-time observables, where one sees
that time-translation
invariance is never achieved, and that there is always a residual dependence
on the age of the system.

Our starting point will be the observation that, even in the presence of
aging, in short-range systems
with non-singular interactions, the long-time limit of the energy
density must coincide with its
equilibrium average.  We then explore the consequences of linear
response theory supposing
equilibration of the energy density, but not necessarily full thermal
equilibrium.  Using some
special random local perturbations of the original Hamiltonian, we
identify a set of
susceptibilities
 which are endowed with very useful properties.  When evaluated at
 equilibrium, these susceptibilities define the moments of
 an order parameter function
 $\tilde P(q)$ which is self-averaging. On the other
 hand, in the dynamical framework, one can
 show that their dynamic expectation values converge at large
 times to their static one, establishing the bridge between
 aging and static
ergodicity breaking.  We will then
establish, for stochastically stable systems,
 the identity of the \textit{fluctuation dissipation ratio}
(FDR), which expresses the
asymptotic violation of FDT, with the overlap probability function that
describes the statistics of
distances among pure states in the equilibrium measure.  This result is
very general, since it holds
in any system with short-range interactions. (In long-range models, or in
models with singular
interactions, the convergence of the energy density to equilibrium is
not guaranteed and the theorem
does not necessarily apply.)  For equilibrium systems it
reduces to the usual relations
connecting static and dynamic susceptibilities, which are consequences
of the FDT. It is
more interesting in the case of glassy systems, where new
consequences can be derived.

Similar arguments can be developed in order
 to study other aspects of ergodicity
breaking.  Particularly
interesting is the  geometrical organization of equilibrium
states in phase space, for
which mean-field theory predicts ultrametricity.  This property has a
dynamical analog in the
ultrametric relation among (two-time) correlation functions for three
different long times.  Indeed,
ultrametricity appears rather naturally in the dynamics within the
``weak ergodicity breaking''
scenario. We shall show that if ultrametricity holds in dynamics, then
stochastic stability implies
ultrametricity for the equilibrium system.

A first partial account of
our results has appeared in
ref.~\cite{noi}.

The paper is organized as follows:
In the next section we recall some basic facts about the out of
equilibrium dynamics and aging. Sect.~\ref{Equilibrium} introduces
the overlap probability function and reviews
 some of the main equilibrium properties. Sect.~\ref{perturb} introduces
 the set of random local perturbations of the Hamiltonian
 which are
 used in order to define the new order parameter function $\tilde P(q)$,
 and its dynamic counterpart. We then develop, in sect.~\ref{stoch_stab},
 the notion of stochastic stability, and show the close link
 which exists between the functions $P(q)$ and $\tilde P(q)$
 in stochastically stable systems. Sect.~\ref{ultrametricity}
 establishes the ultrametricity property of these systems,
 firstly in a dynamical framework, and consequently
 in the statics. Sect.~\ref{conclusion} contains some
 discussion and summary of the results. Two appendices
 explain some technical details of the computations.

\section{Off-equilibrium dynamics}\label{offeq}
  We use for definiteness the language of magnetic systems,
considering classical spins $S_x$
on a $d$-dimensional lattice of size $N=L^d$.  The spins are real
variables in a double-well
potential, and we often implicitly take the Ising limit.  We do not need
to specify much of the
interactions, but we assume that they are expressed by a
\textit{short-range} Hamiltonian
$H(S)$.

We suppose for simplicity that the dynamical
evolution  of the system of $N$ spins is given by the Langevin equation
\begin{equation}
\dot{S_x}=-{\partial H\over\partial S_x}
+\eta_x ,
\end{equation}
where $x$ is a discrete index that labels the spins and $ \eta_x$ is a
white noise of variance
$\langle \eta_x(t)
\eta_y(t')\rangle = 2 T \delta_{xy} \delta(t-t')$.
 We use here the same notation, namely the angular
brackets, to denote both the average over the thermal noise in the dynamic
context, and the Gibbs average in the static context. Which one is meant
should be clear from the context.
The use of the Langevin dynamics  is not crucial for the
statements we shall make below, since they are essentially based on
linear
response theory. We show
 in appendix~\ref{dyn:app} how
our arguments can be
generalized to any system with dissipative dynamics (e.g., Glauber
or Metropolis), where the spins are locally updated according to the
value of the instantaneous molecular field.

 We denote by
${\bf S}$ the microscopic configuration of the system: ${\bf
S}= \{ S_x \} $.  We make use in the following of the Martin-Siggia-Rose
representation of the Langevin dynamics, in which the thermal
averages of the observables $O$ are expressed by functional
integrals:
\begin{equation}
\langle O(t) \rangle =\int {\cal D}({\bf S}) \,
{\cal D}(\hat{\bf S})\; {\rm e}^{I[{\bf S},i\hat{\bf S}]}
O[\bf S(t)],
\end{equation}
with the dynamical action
\begin{equation}
 I[{\bf S},i\hat{\bf S}]= \int d t'\; \sum_x i\hat{S}_x(t')
 \left[\dot{S_x}+ {\partial H\over\partial S_x}
 + iT\hat{S}_x\right].\label{MSR:action}
\end{equation}
When inserted in a correlation function, the auxiliary field $i
\hat{S}_x(t)$ acts as the functional derivative with respect to a
magnetic field at site $x$ and time $t$.

We suppose that the system has been quenched at time $t=0$ from a very
high temperature into its
spin glass phase.  It then evolves at a fixed temperature $T$, starting
from a random initial
configuration.  The fact that the system is off equilibrium appears at
long times in the ``two time
observables'', i.e., in the correlation and response functions.  We
focus on the spin-spin
autocorrelation function $C(t,t')= (1/N) \sum_x \langle S_x(t)
S_x(t')\rangle$, and on the
associated response of the spins at time $t$ to an instantaneous field
at time $t'$: $
R(t,t')=(1/N)\sum_x \delta \langle S_x(t) \rangle /\delta
h_x(t')\equiv (1/N) \sum_x \langle
S_x(t) i\hat{S}_x(t')\rangle$.  In our notations it is understood
that
the first time argument is always larger then the
second one.
A quantity which is of particular
experimental interest is the
relaxation function, which is proportional to the integrated response
function,\footnote {The
definition of $\chi$ adopted here differs by a factor $T$ from the most
common one.}
\begin{equation}
 \chi(t,t')= T \int_{t'}^{t} d t''\, R(t,t'').
\end{equation}
In the linear response regime this function measures
the adimensional magnetization, deriving from a uniform field acting
from time $t'$ to time $t$. We suppose in the following that the
contributions due to the response of a spin to the field in different sites
sum up to zero, otherwise our definitions
must be adapted to take into account this contribution.

We are interested in the behavior of these quantities in the double
long-time limit $t,t'\to
\infty$, always taken after the thermodynamic limit $N\to \infty$.  The
phenomenon of aging is
associated with different behaviors, depending on the path along which
the two times are sent to
infinity\cite{francesi}.  A first regime is obtained when the long-time limit is taken
keeping the difference
$\tau=t-t'$ finite.  In this regime, the correlation function $C(t,t')$
becomes a function of
$\tau$: $c(\tau)$.  The quantity $q_{\rm EA}=\lim_{\tau\to\infty}
c(\tau)$ is known as the
Edwards-Anderson order parameter in spin-glass theory.  In systems which
have only one equilibrium
phase and consequently reach equilibrium very fast, the only non-trivial
long-time regime
corresponds to this limit, and one can measure $q_{\rm EA}$ by taking the
long-time limit of $C(t,t')$ in any order, provided $t-t'\to\infty$.

On the other hand, in systems exhibiting aging there are other
interesting dynamical regimes, where
the correlation relaxes below $q_{\rm EA}$, and spans the whole interval
$[0,q_{\rm EA}]$, depending
on how the infinite-time limit is taken.\footnote{Cases in which the
correlation does not decay to
zero, but to a minimum value $q_0<q_{\rm EA}$ are also possible.  For
simplicity, we always consider
in this paper $q_0=0$.  Considering the generic case would just make
notations heavier.} This
phenomenon has received the name of ``weak ergodicity 
breaking''~\cite{bouchaud}.  It
can be generically expected
whenever there are different phases in competition.  For example, it is
found in the domain-growth
dynamics of ferromagnets, or more generally in spinodal decomposition.
In these cases, if the
initial conditions do not favor one phase with respect to the others,
one finds for long times
$C(t,t')\approx c^*(L(t')/L(t))$, while $\chi(t,t') \approx
\chi_{\infty}+L(t)^{-a}\chi^*(L(t')/L(t))$
\cite{x(q)-sim,fisher}.
 This
dependence can be understood on
the basis of dynamical scaling.  After a short transient the system
enters an asymptotic regime
where the growing domains of the different phases are characterized by a
typical length $L(t)$.
Scale invariance implies for the correlations $C(t,t')\approx
C^*\left(L(t')/L(t)\right) $.
The factor $L(t)^{-a}$
($a$ is equal to 1 in the simplest cases) appearing in $\chi$ is
due to the fact that only
the spins at the boundaries between different domains give a contribution
to the long-time response.
In pure ferromagnets one has $L(t) \sim t^{1/2}$ for non-conserved order
parameter dynamics and
$L(t) \sim t^{1/3}$ for conserved dynamics.  For domain growth in the
presence of some pinning
disorder, the growth of $L(t)$ will be slower, but the general behavior
should be the same as in
pure ferromagnets, with a response function which is much smaller than
the correlation at large
times.

The case of glassy systems, where aging seems to have a very
different nature than in domain-growth processes, is more
interesting.
It is important, in the following discussion, to consider aging in
the linear response function $\chi(t,t')$.  Mean-field theory
predicts that, asymptotically, $C$ and $\chi$ should depend on the
same function of $t$ and $t'$, and therefore, that
$\chi(t,t')\approx \hat \chi\left(C(t,t')\right)$. Such a behavior
cannot take place in domain growth kinetics (nor in a droplet
description of aging), in which only the
spins at the border of the growing domains contribute to the
response.
In order to characterize the relation among correlation and response
one can define the function
\begin{equation}
X(q)=-{\partial \hat \chi \over \partial q}= -
\lim_{{t,t'\to\infty}\atop C({t,t')=q}}
\frac{\partial \chi(t,t') }{\partial t'}\Big/
\frac{\partial C(t,t') }{\partial t'}\equiv
\lim_{{t,t'\to\infty}\atop C({t,t')=q}}
T R(t,t')\Big/
\frac{\partial C(t,t') }{\partial t'},
\label{fdrat}
\end{equation}
which is called the fluctuation-dissipation ratio, and depends a priori
on the value $q$ of the
correlation which is kept fixed while one sends the two times $t,t'$ to
infinity.

In the finite $\tau$ regime, in which $q\in [q_{\rm EA},1]$, local
equilibrium implies $X(q)=1$,
which is the expression of the usual fluctuation-dissipation theorem.
But a non-trivial behavior is
possible for $q\in [0,q_{\rm EA}]$.  Mean-field theory predicts that the
function $dX/dq$ is non
trivial, and satisfies the properties of a probability distribution.
Simulations have shown that
this fact also holds in 3D and 4D spin glasses \cite{x(q)-sim}.  The
function $X(q)$ is experimentally accessible
through waiting-time monitored measurements of the thermoremanent
magnetization and
of the noise.  Its measurement will allow one to distinguish whether the
glassy behavior is a slow
domain growth, where $X(q)=\theta(q-q_{\rm EA})$, or
whether it has a more
complicated behavior, with $X(q) \ne 0$ at $q<q_{\rm EA}$, as seen in
mean field and in simulations.
%\marginpar{Crr. 16.3.99}

\section{Equilibrium}
\label{Equilibrium}

For short-range Hamiltonians with non singular interactions, classical
arguments, based on neglecting
surface with respect to volume terms, show that the averages  of the
free energy density have a
thermodynamic limit.
Intensive quantities, which can be obtained as derivatives of the free
energy densities are unique
and self averaging except, in case, at first order
phase transitions points, where more than
one phase is present.  We
are interested in situations of broken ergodicity, where the
infinite-volume equilibrium measure is
not unique.  Sometimes, when one knows a priori the set of
equilibrium states, one can
select one element of the
set of the equilibrium measures or another
by properly
choosing the boundary conditions or some small external field.
In ordered systems this
allows to select \textit{pure
phases}, or ergodic components, which correspond to physical equilibrium
states.  Unfortunately,
this procedure is of no use in the case of glassy systems, where our
lack of knowledge on the
structure of the low lying states prevents us from picking up the
correct boundary conditions
necessary to project on the pure phases.
Indeed, pure phases have identical values of the densities of
extensive quantities like internal energy, magnetization etc.

If the boundary conditions are
uncorrelated with the
energy landscape (e.g., periodic) the Gibbs measure turns out to be a
mixed measure.  In addition
the measures can have a strong dependence on the quenched disorder
and/or on the system size.
Under these conditions it is useful to study the statistical properties
of the mixtures and to
perform the analysis at very large, but finite $N$, avoiding the
discussion of
pure states for an actually infinite system.  An order
parameter function which shows up naturally in mean field is the overlap
probability function.
 It is defined by picking up  two
microscopic configurations
${\bf S}$ and ${\bf S'}$ at random, with the Gibbs measure, and
measuring the histogram $P(q)$ of
their overlap $ Q({\bf S},{\bf S'})=(1/N)\sum_x S_x S'_x $.  More
precisely the order
parameter function, $P(q)$, is given by
\begin{equation}
P(q) =\lim_{N\to\infty} E_J \left[\frac {1}{Z^2} \sum_{{\bf S},{\bf
S'}}\exp(-\beta
(H({\bf S})+H({\bf S'}))) \,\delta \left(Q({\bf S},{\bf S'})
-q\right)\right]. \label{pidiqu}
\end{equation}
We have denoted by $E_J$ the average over the quenched disorder in the
system (if present) or over a
window in $N$, increasing with $N$,
and by $Z$ the partition function.
In ergodic systems, $P(q)$ is
given by a single delta function, whereas it can have a nontrivial
structure in systems with
ergodicity breaking.  This appears explicitly in several mean-field
models.  Present numerical
simulations give evidence in favor of ergodicity breaking and of a
non-trivial $P(q)$ in
finite-dimensional spin-glass models \cite{romani-young}, but the
relevance of this result has been
challenged\cite{NS2}.  The theorem we are going to discuss
relates the function $P(q)$, which
is not directly accessible in experiments, to the fluctuation
dissipation ratio $X(q)$, which is
measurable.  It thus clarifies the relevance of the function $P(q)$
defined by eq.~(\ref{pidiqu}) and opens the way to
obtain some experimental evidence concerning its behavior.

In the next section we shall need to compute some correlation functions
like
\begin{equation}
{\cal C}_p=E_J \left[\langle S_{x_1}\cdots S_{x_p} \rangle^2 \right]=
E_J\left[\frac {1}{Z^2}
\sum_{\bf S,S'} \exp(-\beta (H({\bf S})+H({\bf S'}))\,S_{x_1}
S_{x_1}'\cdots S_{x_p}S_{x_p}'\right], \label{qk}
\end{equation}
for very far apart sites $x_1,\ldots,x_p$, in the presence of some
quenched random variables $J$.  The
relation of these moments with physically clustering states and its
expression in terms of the
function $P(q)$ is discussed in \cite{MPV}. We just give here the result:
\begin{equation}
 {\cal C}_p=\int dq \; P(q)\,q^p.
\end{equation}

\section{ Perturbing the Hamiltonian: a new order parameter function}
\label{perturb}
In this section we shall define a new equilibrium order
parameter function $\tilde P(q)$, and the corresponding
fluctuation-dissipation ratio
$\tilde X(q)$.  The derivation of their properties
involves the study of the linear response to some special sets of
perturbations of the original
Hamiltonian.  This method has been recently used to derive interesting 
properties
of the overlap distribution at equilibrium \cite{guerra,aizenman}.
We start by recalling some well-known
perturbations which are long
ranged, and whose expectation values yield the moments of the
function $P(q)$ in statics and
of $dX/dq$ in dynamics.  However, the long-range nature of the
perturbation forbids one to establish the
relationship between the long time dynamics and the statics.  We shall
thus introduce a new set of
perturbations which are purely local, and from which the result can be
derived.

It is known that the moments of $P(q)$ and $dX/dq$ are respectively
related to the canonical
\cite{guerra} and the dynamical averages \cite{cuku} of long-range
perturbations to the Hamiltonian
of the form
\begin{equation}
H_p^{\rm LR}(S)=\sum_{i_1<\cdots<i_p}^{1,N}
J_{i_1,\ldots,i_p}S_{i_1}\cdots S_{i_p}, \label{HLR}
\end{equation}
where the couplings $J_{i_1,\ldots,i_p}$ are independent Gaussian
variables with zero mean and
variance $E_J(J_{i_1,\ldots,i_p}^2)=p!/(2 N^{p-1})$.
Now, one can easily see that the
canonical average of $H_p^{\rm LR}$
with a perturbed Hamiltonian given by
\begin{equation}
H_\epsilon=H_0+\epsilon H_p^{\rm LR},
\end{equation}
verifies, for all values of $\epsilon$,
\begin{equation}
E_J \langle H_p^{\rm LR} \rangle = -\beta \epsilon N \left( 1-\int
dq \; P_\epsilon(q)\, q^p \right), \label{SLR}
\end{equation}
irrespective of the specific form of $H_0$. Here the function
$P_\epsilon(q)$ is the order parameter function (\ref{pidiqu}), in the
presence of the perturbing term in the Hamiltonian.
The derivation, reported in appendix ~\ref{dyn:app}, involves
an integration by parts in a finite system, followed by the infinite-volume
limit.

On the other hand, in off-equilibrium dynamics, one has
\begin{equation}
 \lim_{t\to\infty} E_J \langle H_p^{\rm LR}(t)
\rangle
 = -
 N\beta \eps \left( 1-\int dq \; \frac{d X_\epsilon(q)}{dq}\,
q^p
\right), \label{DLR}
\end{equation}
where $X_\epsilon(q)$ is the FDR in the presence of the
perturbation (we recall that the system has been quenched at time zero). For
completeness, formulae
(\ref{SLR}) and (\ref{DLR}) are derived in appendix~\ref{dyn:app}.

These identities suggest that one could have $P(q)=dX/dq$.
However, in order to derive this
result one needs two more steps which are far from trivial.  First, one
should prove that $H_p^{\rm
LR}$ is statically self-averaging and that its dynamic expectation value
tends to the equilibrium
one.  Second, one must discuss the continuity of the functions at $\eps
=0$.  Concerning the first
point, our main concern is that the long-range perturbation could give
rise to infinitely long-lived
metastable states in the system, so that $\lim_{t\to\infty}
\langle H_p^{\rm
LR}(t)\rangle_{\rm dyn} \ne
\langle H_p^{\rm LR}\rangle_{\rm stat}$. Although we feel that
this is not likely to happen, in most systems, for small enough
$\epsilon$,
we do not know at present how to prove this fact.
Below we use local (short-range) perturbations which get
around this problem, and in
the next section we address the issue of continuity at small
$\eps$.

The strategy we follow amounts to modify the definition of the
perturbations, introducing a
new $H_p$ which is an extensive sum of some local observables.  Then the
whole Hamiltonian $H_0+
\eps H_p$ is short range, and thus self-averaging, and the
long-time equilibration of the
dynamical expectation $\left<H_p(t)\right>$
towards its equilibrium value can be
shown as follows.  One starts by
proving that the free energy equilibrates.  The classical proof of this
fact is based on nucleation
arguments.  One first notices that the time-dependent free-energy
density must reach, for long
times, the equilibrium value $f_{\rm eq}$.  All metastable states with a
higher free-energy density
$f^*$ can be destabilized by the nucleation of a bubble of radius $r$
with a free energy cost equal
or lower than $c r^{d-1} + (f_{\rm eq}-f^*) r^d$.  Therefore the
free energy density reaches
its equilibrium value.  Since the static perturbation energy $ \left< 
H_p\right>$
is the derivative of $\beta F $ with respect
to $\eps$, the equilibration of the free energy also implies the
equilibration of $ \left< H_p(t)\right>$.\footnote{Generically,
one should qualify this statement at first-order phase transition points,
where the perturbation can select a given phase. However,
as previously noticed, we are interested in a situation in which
all pure phases have the same values of extensive observables, and this
does not happen.}  This
differentiation will be allowed at
small $\eps$ provided the free energy is regular enough at small $\eps$,
which is the issue which we
shall discuss in the next section.

The local perturbing Hamiltonian can be chosen in several ways,
 we present here the one we already considered in \cite{noi}.
Let us consider a set of operators which translate the lattice
along a direction $\hat{\imath}$ parallel to one of the coordinate
axes:
\begin{equation}
{\cal T}_k^{(p)}(x)=x+\frac{k L}{p} \hat{\imath};\qquad
k=1,\ldots,p-1.
\end{equation}
We denote by ${\cal S}_1$ the set in which the coordinate in the
direction $\hat{\imath}$ takes the values $1,2,\ldots,L/p$. The
new perturbations will have the form
\begin{equation}
H_p({\bf S})= \sum_{x\in {\cal S}_1} J_x^{(p)} S_x S_{{\cal
T}_1^{(p)}(x)} \cdots S_{{\cal T}_{p-1}^{(p)}(x)}, \label{pert-sr}
\end{equation}
where the couplings $J_x^{(p)}$ are independent, identically
distributed Gaussian variables, with zero mean and variance $E_J
J_x^2=p$.
At first sight, also these perturbations seem to contain long-range
interactions.  However,  they can be viewed as
short-range observables in a different space.  This is more easily
illustrated for $p=2$, but the generalization to arbitrary $p$ is
immediate.
For $p=2$ we can divide the space into two halves (${\cal S}_{l}$
and ${\cal S}_r$) and rename the spins in the right-hand half so
that if $x \in {\cal S}_l$ then ${\cal T}(x) \in {\cal S}_r$ and
$S_{{\cal T}(x)}=S'_x$ (we wrote ${\cal T}(x)$ for ${{\cal
T}_1^{(2)}(x)}$).  The total Hamiltonian can now be written
as
\begin{equation}
H({\bf S},{\bf S'})=H_l({\bf  S})+H_r({\bf  S'})+B({\bf S},{\bf
S'}) +\epsilon\sum_{x\in {\cal S}_l} J^{(2)}_x S_x S_x'.
\label{hloc}
\end{equation}
The Hamiltonians $H_l$ and $H_r$ refer respectively to the spins in
${\cal S}_{l}$ and ${\cal S}_r$.
The term $B({\bf S},{\bf S'})$ is a surface term whose presence does not
affect the average of
$H_2$.  Dropping it, the Hamiltonian (\ref{hloc}) characterizes a spin
system of size $L^{d}/2$,
with two spins $S_x,S_x'$ on each site, and a purely local interaction.
The static expectation
value of the perturbation $\langle H_2 \rangle$, gives therefore a
contribution to the internal
energy of the system which is extensive and self-averaging, i.e.,
independent (in the
thermodynamical limit) of the particular realization of the disorder
contained in either $H_0$ or
$H_2$.  Its long-time limit in off-equilibrium dynamics must yield its
canonical value.

We now
take advantage of the self-averaging property of $H_2$ to express its
static and dynamic values as
functions of $P(q)$ and $X(q)$ respectively.  Let us first discuss the
statics:
\begin{eqnarray}
 \langle H_2\rangle & = & E_J\, \langle H_2 \rangle\nonumber\\
& = & E_J \left[  \int {\cal D}({\bf S}) \, \exp\left( -\beta (H({\bf
S})+\epsilon H_2({\bf S})) \right) \,\sum_x J^{(2)}_x S_{x} S_{{\cal
T}(x)} \right].
\end{eqnarray}

Integrating by parts over the random couplings $J^{(2)}_x$ we find
\begin{equation}
 \langle H_2 \rangle=\beta\epsilon N (1-E_J \langle S_x S_{{\cal T}(x)}
\rangle^2).
\end{equation}
In the linear response regime we can write
\begin{equation}
E_J \langle S_x S_{{\cal T}(x)} \rangle^2 =E_J \langle S_x S_y
\rangle^2+ O(\epsilon),
\end{equation}
where $x$ and $y$ are far away sites not directly coupled in the
Hamiltonian, and consequently, according to the discussion of the
previous section,
\begin{equation}
\langle H_2 \rangle=\beta\epsilon N \left(1-\int d q\;
P_{\epsilon}(q) \,q^2\right), \label{estat} \label{q2}
\end{equation}
where the function $P_{\epsilon}(q)$ is the average overlap
probability function of the perturbed system. Let us remark that
in order to derive (\ref{q2}) we need to integrate over the couplings
$J^{(2)}_x$. In the case where also $H_0$ contains quenched disorder we
can choose to integrate over it or not, but the result should be
independent of this operation. This tells us that the order parameter
function, averaged over the couplings in $H_2$, is
self-averaging with respect to the quenched variables in $H_0$,
 even for infinitesimal $\epsilon$.

Similar considerations hold in the case of the dynamics. The
self-averaging property of $H_2$ in the dynamics allows us to
write
\begin{eqnarray}
 \langle H_2 (t) \rangle & = & E_J\, \langle H_2(t)\rangle\nonumber\\
& = & E_J \int {\cal D}({\bf S}) \, {\cal D}(\hat{\bf S})\; {\rm
e}^{I[{\bf S},i\hat{\bf S}]}\, { \sum_x   J^{(2)}_x S_{x}(t) S_{{\cal
T}(x)}(t)} .
\end{eqnarray}
Integrating by parts over the $J^{(2)}_x$'s, and observing that the
insertion of $i\hat{S}_x(t')$ acts as the derivative with respect
to an impulsive magnetic field at site $x$ and at time $t'$ ($i\hat{S}_x(t')
\longmapsto\delta/\delta h_x(t')$) we obtain
\begin{equation}
 E_J\, \langle H_2(t)\rangle=2\epsilon\sum_x
E_J \left[ \int_0^t dt' \;
\frac{\delta}{\delta h_{{\cal T}(x)}(t')}
\langle S_x(t) S_x(t') S_{{\cal T}(x)}(t)\rangle\right].
\end{equation}
In the linear response regime, $\beta \epsilon\ll 1$, the average
of the product on far-away sites factorizes up to terms of order
$\epsilon$, and one has
\begin{equation}
E_J \left[\frac{\delta}{\delta h_{{\cal T}(x)}(t')}\langle S_x(t)
S_x(t') S_{{\cal T}(x)}(t)\rangle\right] =C(t,t') R(t,t') +O(\epsilon).
\label{cr}
\end{equation}
Assuming that the bound holds uniformly in time and substituting
the definition (\ref{fdrat}) of the {FDR}, we obtain, for large
values of $t$,
\begin{equation}
2 \epsilon\beta N \int_{0}^1 dq\; X_\epsilon(q)\,  q= \epsilon \beta N\left(
1
-\int_{0}^1 dq \; {dX_\epsilon \over dq} \, q^2\right),
\label{edyn}
\end{equation}
where we have used $C(t,t)=1$, and $\lim_{t \to \infty} C(t,0)=0$
(see footnote 2).  We have denoted by $X_\epsilon$ the {FDR} of the
system with the perturbed Hamiltonian $H_\epsilon$.  Comparing the two
results, (\ref{edyn}) for the dynamics and (\ref{estat}) for the
statics, we see that the second moments of the dynamical order
parameter function $dX_\epsilon(q)/dq$ and of the static one
$P_\epsilon(q)$ coincide for the system in the presence of the
perturbation $\epsilon H_2$.  It is straightforward to generalize this
derivation to arbitrary $p$ (for $p=1$ the perturbation is nothing but
a small random field term), which shows that the $p$-th moments of the
two functions $dX_\epsilon(q)/dq$ and $P_\epsilon(q)$ coincide for
small $\epsilon$.  Therefore the two functions coincide, and their
limits for $\epsilon\to 0$, $d \tilde{X}(q)/d q$ and $\tilde{P}(q)$
respectively, also coincide.
 We have thus
found one ``new'' order parameter function which describes both
the static equilibrium situation and the out of
equilibrium aging dynamics. In the next section we will
see to what extent one can relate these functions to the FDR and
order parameter function of the unperturbed system.

Finally let us point out that the choice of the perturbation is by no means
unique.
In principle we could also couple different systems instead of
translated copies of the same systems.
This would amount to introducing $p$ different systems and  using as as
perturbation
\begin{equation}
H_p({\bf S})= \sum_{x} J_x^{(p)} S_x^{1}S_x^{2} \cdots  S_x^{p},
\label{pert-diff}
\end{equation}
where the unperturbed Hamiltonian is given by the sum of
Hamiltonians of the $p$ uncoupled system.  This approach is also
viable and yields the same results as the one discussed above.
Another possibility is to consider just one system with
perturbing interactions with a finite range (Kac potential type), to be
sent to infinity after the thermodynamic limit.

Let us conclude by noticing
\textit{en passant} that we can use
the short-range perturbations (\ref{pert-sr}) to repeat the derivation
by Ghirlanda and
Guerra~\cite{GG} of the identities for the probability function of
two overlaps at equilibrium,
avoiding also in this case problems related with the proof of
self-averaging and continuity of the
correlation functions for long-range perturbations.

\section{Stochastic stability}
\label{stoch_stab}
We have seen in the last section that the static and dynamic order
parameter distribution functions,
respectively $P_\epsilon(q)$ and $dX_\epsilon/dq$, are equal in the
presence of the perturbation,
and that therefore their $\epsilon\to 0$ limits, $\tilde P(q)$ and
$d\tilde X/dq$, coincide.  We now
address the question of the relation between these quantities and
the conventional order
parameter functions $P(q)$ and $dX/dq$, i.e., the order parameter
function and FDR defined in the
absence of the perturbation.  If we were to apply naive perturbation
theory, we would directly
conclude that they are equal.  However, while this conclusion is
inescapable for ergodic systems, it
needs to be qualified in the more interesting case in which many ergodic
components are present.

Let us start with the statics.
The problem is to understand what is the ``equilibrium'' state in the
presence of the perturbation
$\eps H_p$, and how it is related to the equilibrium state at $\eps=0$.
Some problems can arise
whenever the equilibrium expectation value of $H_p$ is not the same for
all pure phases of the
unperturbed systems.  Then the limit value of $H_p$ as $\epsilon\to 0$
will be the one corresponding
to the favored phases.  A simple example is the Ising model in the
ferromagnetic phase with free
boundary conditions, to which one adds a negative magnetic field term as
a perturbation.  Then in
the vanishing field limit, the long-time limit of the perturbation is
evaluated in the state with
negative spontaneous magnetization, while the unperturbed measure
corresponds to a mixture of the
positive and negative magnetization pure states.  Clearly in this case
$\tilde P(q) \ne P(q)$.
These two functions will also differ if the perturbation, instead of
being a uniform field, is a
random field.

Indeed, the stochastic perturbations that we have considered will in
general reshuffle the weights
of the different ergodic components in the Gibbs measure, or even change
their nature, and this
changes the $P(q)$ function to a different one ($\tilde P (q)$).
However our perturbations are
random perturbations which are not correlated with the original
Hamiltonian.  So they should change
the free energies of the various pure states of the original systems by
random amounts.  Usually in
glassy systems the \textit{distribution} of these free energies is stable
under independent random
increments, as has been shown in mean field
(in fact this
property lies at the heart of the cavity method \cite{MPV}).
If this is the case then
the two functions $\tilde
P(q)$ and $P(q)$ will coincide.  A noticeable exception to this result is the
case where the original
Hamiltonian has an exact symmetry, which is lifted by the perturbation.
The simplest case is that
of a spin glass with a Hamiltonian invariant under spin inversion.  In
this case $P(q)=P(-q)$, since
each pure state appears with the same weight as its opposite in the
unperturbed Gibbs measure.  On
the other hand, if we consider $H_p$ with odd $p$, this symmetry is
lifted.  This means that in the
$\eps \to 0$ limit only half of the states are kept.  If the reshuffling
of their free energies is
indeed random, then we shall have $\tilde P(q)=2\theta(q)P(q)
\equiv \hat P(q)$.  The same type of
reasoning applies whenever the overlap $q$ transforms according to a
representation of the symmetry
group of the unperturbed Hamiltonian $H_0$.

Let us now discuss the case of dynamics. It is clear that
the finite-time response and correlation functions involved in
the definition of the FDR are continuous
functions of $\eps$ for $\eps\to 0$. If this limit
is uniform in time, so that the infinite time and the $\eps\to 0$ limits
commute, one has $\tilde X (q)=X(q)$. On the other hand,
it may be the case that the linear response regime shrinks
to zero as the time goes to infinity.
This possibility shows up in the situation, in which the
perturbations favor one (or more phases) phase.
However we stick here to random perturbations such that
the expectations of $H_p$ vanishes at $\eps=0$. In this
case it is reasonable to assume that the linear response
regime survives at very long times, implying that
$\tilde{X}(q)=X(q)$. Let us stress that the existence of
a linear response regime uniform in time is implicitly
assumed in experiments attempting to measure $X(q)$ in
real systems, and is anyway a question susceptible of experimental 
investigation.

Once the effect of exact symmetries is taken into account, one may
expect that, for a large class
of systems, the limit function $\tilde P(q)$ in the limit of small
perturbations tends
to the order parameter function $\hat P(q)$ of the pure system
where the exact symmetries are
lifted, what is nothing but a statement of continuity of the
correlations at small $\eps$.  We call
this continuity property of the correlations \textit{stochastic
stability}.  Ordinary systems with
symmetry breaking and mean-field spin glasses are examples of
stochastically stable systems.
In symmetry breaking systems (and in ergodic systems),
the equality of $\tilde{P}$ and $\hat{P}$ is
immediate, since both functions consist in a single delta
function. Thus, the problem of deriving the equality between $\tilde{P}$ and
$\hat{P}$, appears only when the coexisting phases are unrelated
by symmetry.

Unfortunately, we are not able to characterize the class of
stochastically stable systems in
general.  In particular, we do not know for sure whether short-range
spin glass, for which our
theorem is most interesting, belong to this class.  However, stochastic
stability has been
established rigorously in mean field problems \cite{guerra,aizenman}, and
numerically exhibited in three and four dimensional spin glasses, where
also the equality between
$P(q)$ and ${dX}/{dq}$ has been confirmed ~\cite{ters}.

Stochastic stability
is a very powerful property,
and  it is the ingredient which allows to
relate the properties of the low lying configurations, which
dominate the Gibbs measure, to those of the configurations much higher in
energy which are seen in the dynamics.
This is most easily explained in the usual framework of replica-symmetry
breaking, considering an approximation
with only two possible values of the overlap, $q_{0}=0$ among different states
and $q_{1}$ among the same
state (i.e.,  one-step replica-symmetry breaking).
The probability of
finding a state with total
free energy $F_{\al}=F$ is given by $\rho(F-F_0)$,
where $F_0$ is the equilibrium free energy.  The weight of each
state is given by
\be
w_{\al}\propto \exp (-\beta F_{\al}).\ee
In one-step replica-symmetry breaking,
  the states which contribute
to the Gibbs measure have  nearly degenerate
 free energies. The non-extensive
fluctuation of their free energies, corresponding to the low $F$ regime
of $\rho(F)$, is given by \cite{grossmez,mpv_free}
\be
\rho(F)\propto \exp(\beta m F),\label{ONESTEP}
\ee
and the function $P(q)$ is given by
\be
P(q)=m \delta(q-q_{0})+(1-m) \delta(q-q_{1})\ .
\ee
In its dynamical evolution from a random initial state,
the
total energy of a configuration at time $t$
decreases as
\be
E=E_{\infty} +cN t^{-\lambda}\ ,
\ee
where  $\lambda$ is an appropriate exponent. Correspondingly the difference
between the total free energy at time $t$ and the equilibrium value will be
always of order $N$, with a prefactor going to zero when $t$ goes to
infinity. One  can physically argue that the dynamics only probes
the
behavior of the function $\rho(F)$ at large argument, and should not be
related to the statics.
Stochastic stability solves this apparent paradox, because
it forces the
function $\rho(F)$ to be of the form (\ref{ONESTEP}),
not only when $F-F_0$ is finite, but
also in the range where $F-F_0$ is \textit{extensive} but small
(say of order $\eps N$).
Indeed it imposes
that the form of the function
$\rho(F)$ remains unchanged  (apart
from a possible shift in
$F_{0}$) when one adds a small random perturbation.
Consider the effect of a perturbation of strength
$\epsilon$ on the free
energy of a state, say $\alpha$. The unperturbed
value of the free energy is denoted by $F_\alpha$. The  new
value of the free energy $G_{\al}$ is given by
$
G_{\al}=F_{\al}+\eps r_{\al}
$
where $ r_{\al}$ are identically distributed
 uncorrelated random numbers.
Stochastic stability implies that the distribution
$\rho(G)$ is the same as $\rho(F)$. Expanding to
second order in $\epsilon$ we see that this implies
$d\rho/dF\propto d^2\rho/dF^2$,
whose only physical solution (apart the trivial
one  $\rho(F)=0$, which corresponds to non-glassy systems)
is given by eq.~(\ref{ONESTEP}).
The same conclusion could be obtained using the methods of reference
\cite{PARI} computing the sample-to-sample fluctuations
of the function $P_{J}(q)$, which in this case, where ultrametricity is
trivially satisfied, are
completely determined by the knowledge of of the function $P(q)$.
We  see that stochastic stability fixes the form of the
function $\rho$ and therefore
connects in an inextricable way the low and the high free energy part
of the function $\rho$, avoiding
a possible paradox.

\section{Ultrametricity and separability}
\label{ultrametricity}
A very interesting aspect of mean-field spin glasses at equilibrium is
the hierarchical organization
of the pure phases, which build an ultrametric space.
Indeed, in these models, the probability
distribution of three overlaps
\begin{eqnarray}
P^{(3)}(q_{12},q_{13},q_{23})& = & E_J \frac {1}{Z^3} \sum_{{\bf
S},{\bf S'},{\bf S''}}\, \exp(-\beta (H({\bf S})+H({\bf
S'})+H({\bf S''}))) \nonumber\\  &&\qquad{}\times\delta ( Q({\bf
S},{\bf S'}) -q_{12}) \delta (Q({\bf S},{\bf S''}) -q_{13})\,
\delta ( Q({\bf S'},{\bf S''}) -q_{23}), \label{p3}
\end{eqnarray}
has been shown to vanish unless its arguments verify the
\textit{ultrametric inequality}
\begin{equation}
q_{12}\ge \min \{ q_{13},q_{23} \}.
\end{equation}

It would be extremely interesting to measure the function
$P^{(3)}(q_{12},q_{13},q_{23})$ in real
materials.  However, it is clear that a direct measure is impossible for
the same fundamental
reasons as the direct measure of $P(q)$ is impossible: impossibility to
reach equilibrium,
impossibility to access the microscopic structure of the pure phases.

In the following we first review the dynamical analog of the
ultrametric identity.
We then provide some arguments in favor of a link between
the dynamic ultrametricity and the ultrametricity in the
static organization of pure phases, as well as
some arguments which indicate
that dynamical ultrametricity could be a generic property
of stochastically stable systems.
 The arguments are of the same nature as the ones used in the
 previous section, in that they rely
 on the study of systems weakly perturbed
 by some appropriately chosen random couplings. We need to introduce
 additional assumptions, which make the
 conclusion more conjectural.

In glassy dynamics, weak ergodicity breaking scenario implies more
 than just the decay of the
autocorrelation at long times.  In particular, it implies a
relation among the three
different two-time
correlation functions that can be built from the configurations of the
systems at three different
(long) times $t_{\rm min}$, $t_{\rm int}$, $t_{\rm max}$, namely
$C(t_{\rm int},t_{\rm min})$,
$C(t_{\rm max},t_{\rm min})$, $C(t_{\rm max},t_{\rm int})$.  Exploiting
the monotonicity of $C$ in
both time arguments one can, for each $t_{\rm int}$, invert the relation
between $C(t_{\rm
int},t_{\rm min})$ and $t_{\rm min}$ and the one between $C(t_{\rm
max},t_{\rm int})$ and $t_{\rm
max}$.  In this way, for each $t_{\rm int}$, $C(t_{\rm max},t_{\rm
min})$ can be expressed as a
function of $C(t_{\rm int},t_{\rm min})$ and $C(t_{\rm max},t_{\rm
int})$, yielding \cite{cukusk}
\begin{equation}
C(t_{\rm max},t_{\rm min})=f_{t_{\rm int}}\left(C(t_{\rm
int},t_{\rm min}),C(t_{\rm max},t_{\rm int})\right).
\end{equation}
In the long-time limit, keeping fixed $q_{12}=C(t_{\rm int},t_{\rm
min})$ and $q_{23}=C(t_{\rm
max},t_{\rm int})$, the function $f_{t_{\rm int}}(q_{12},q_{23})$ tends
to a limit function
$f(q_{12},q_{23})$.  The properties of this function have been
studied in great detail
in~\cite{cukusk}.  The ``fixed points'' of $f$, defined as the values of
$q$ for which $q=f(q,q)$
are especially relevant.  Indeed, if $q$ and $q'$ are fixed points, then
$f$ verifies the
ultrametric relation $f(q,q')=\min\{q,q'\}$.  The scaling form $C(t,t')=
c^*(h(t')/h(t))$, where
$h(t)$ is some monotonically increasing function of $t$, holds for
values of the correlation between
contiguous fixed points.  This shows that in the limit where the two
times $t,t'$ become large, the
corresponding two-time plane can be divided into sectors, called ${\cal
D}_u$, which we label by
the index $u$.  Each sector ${\cal D}_u$ is characterized by a
monotonically increasing function of
$t$, $h_u(t)$, and is defined by sending the two times $t,t'$ to
infinity with a fixed ratio
$\lambda=h_u(t')/h_u(t)$. One then has $C(t,t') \to
c^*_u(\lambda)$.  The various sectors
have a hierarchical organization \cite{FM} in the sense that, if $t,t' \in {\cal
D}_u$ and $t',t'' \in {\cal
D}_v$, then $t,t'' \in {\cal D}_{\min(u,v)}$.

So far this structure is just one way of expressing the long time
limit of the function of two variables $C(t,t')$, in terms of a
(possibly continuous) set of functions of one variable
$c^*_u(\lambda)$.  It is a very general structure which relies
only on weak ergodicity breaking.  However this structure becomes
much more interesting in view of the explicit solution of specific
mean field models, which has shown that the fluctuation-dissipation
ratio $X(q)$ is constant in each domain ${\cal D}_u$, and
correspondingly,
\begin{equation}
\lim_{{t,t'\to\infty}\atop{h_u(t)/h_u(t')=\lambda}}
TR(t,t')\Big/\frac{\partial C(t,t')}{\partial t'}=X_u,
\end{equation}
independently of the value of $\lambda$.
  This really implies a reduction
of the number of variables, since it means that the integrated
response is just proportional to the correlation in each domain.  We
will see that this hypothesis has far reaching consequences.
 We shall define
the property of dynamical ultrametricity as the fact of having both
the hierarchical domain organization, together with a constant
fluctuation dissipation ratio within each domain.

We shall now assume that we have a stochastically stable systems.  Using
our general strategy of
linear response to random perturbations, we first argue that if it
is dynamically ultrametric, then it is
very likely to be ultrametric in the sense of the equilibrium
distribution.  We then relate dynamical ultrametricity to
the property of \textit{separability}, introduced in \cite{PARI}, which states
that pairs of states with a given overlap $q$ cannot be distinguished
by the value of any differently defined generalized overlap.  For
example we can define an ``energy overlap" of two configurations as
$Q^{\rm H}({\bf S},{\bf S'})=(1/N) \sum_x h_x({\bf S}) h_x({\bf S'})$,
where
$h_x(\cdot)$ denotes the molecular field on site $x$ corresponding to
a given spin configuration. Separability means that, for any pair
$(\alpha,\beta)$ of pure phases, whose overlap $q_{\alpha\beta}$
is equal to $q$, $q^{\rm H}$ is a self-averaging quantity $q^{\rm H}=f(q)$
that depends only on $q$.
A trivial example
of non-separable system, which is neither stochastically stable, is
given by the union of two separable systems. Indeed it is clear that
\begin{equation}
q_{\rm tot}^{\rm H} =q_{1}^{\rm H}+q_{2}^{\rm H}=f(q_1)+f(q_2)\ne
f(q_1+q_2),
\end{equation}
as long as $f$ is non-linear.
Separability is therefore a strong requirement in the sense that
it implies strong correlations among different parts of a system.

We now generalize the static-dynamic equalities derived above to the
three-point probability
distribution function $P^{(3)}(q_{12},q_{13},q_{23})$, relevant for the
discussion of the
ultrametric organization of pure phases.  As in the previous case, we
must identify a set of
susceptibilities generated by this functional.  Using the same
construction as in
section~\ref{perturb}, let us divide the lattice in $p=l+m+n$ slices and
consider two perturbations
of the kind
\begin{equation}
H_p^{(1)}({\bf S})= \sum_{x\in {\cal S}_1} J_x S_x S_{{\cal
T}_1^{(p)}(x)} S_{{\cal
T}_2^{(p)}(x)} \cdots S_{{\cal T}_{l+m-1}^{(p)}(x)},
\label{pert-1}
\end{equation}
and
\begin{equation}
H_p^{(2)}({\bf S})= \sum_{x\in {\cal S}_1} J'_x S_{{\cal
T}_{l}^{(p)}(x)} S_{{\cal
T}_{l+1}^{(p)}(x)}\cdots S_{{\cal T}_{p-1}^{(p)}(x)},
\label{pert-2}
\end{equation}
with \textit{two} different realizations $J_x$ and $J'_x$ of the
quenched couplings.  Notice that
$H_p^{(1)}$ couples spins belonging to the first $l+m$ slices, while
$H_p^{(2)}$ couples spins
belonging to the last $m+n$ slices.  The total Hamiltonian is now
$H_{\epsilon\epsilon'}=H_0+\epsilon H_p^{(1)}+\epsilon' H_p^{(2)}$.  We
now evaluate the average of
the observable
\begin{equation}
O({\bf S})= \sum_{x\in {\cal S}_1} J_x J'_x S_x S_{{\cal
T}_1^{(p)}(x)} \cdots S_{{\cal T}_{l-1}^{(p)}(x)}\times S_{{\cal
T}_{l+m}^{(p)}(x)} \cdots S_{{\cal T}_{p-1}^{(p)}(x)}, \label{obs}
\end{equation}
where the spins in the first $l$ slices are coupled with the spins of
the last $n$ slices.  As in
the case of the observables $H_p$, useful expressions for the dynamic
and the static averages of $O$
are obtained by exploiting the self-averaging property and by
integrating by parts over the variables
$J_x$ and $J'_x$.  The calculations are lengthy, but conceptually
similar to the ones relative to
the observables $H_p$.  They are reported in Appendix~\ref{ultr:app}.
We thus obtain the following
expression of the static average:
\begin{eqnarray}
\langle O\rangle &=& \beta^2 \epsilon \epsilon'
\frac{N}{p} \left( 1-\int_0^1 dq \; P_\epsilon (q)
(q^{l+m}+q^{l+n}+q^{n+m})\right.\nonumber\\ &&\phantom{\beta^2
\epsilon \epsilon' \frac{N}{p}\big(}\left.{} + 2 \int_0^1
dq_{12}\, dq_{13}\, dq_{23}\; P^{(3)}_\epsilon
(q_{12},q_{13},q_{23})\, q_{12}^n q_{13}^m q_{23}^l
\right). \label{stat1}
\end{eqnarray}

Correspondingly, in off-equilibrium dynamics, we find the following
expression of
$\left<O(t)\right>$ in terms of the time-dependent correlation and
response functions:
\begin{eqnarray}
E_J E_{J'} \left<{O}(t) \right> &=& \beta^2 \epsilon \epsilon'
\frac{N}{p} \int_0^t dv \int_0^{t} du
\nonumber\\
 &&{}\times\left(
\left[l C(t,v)^{l-1} R(t,v)\right] \left[ C(v,u)^m\right] \left[ n
C(t,u)^{n-1} R(t,u)\right] \right.
 \nonumber\\
 &&\phantom{ \times
(}{}+ \left[l C(t,v)^{l-1} R(t,v)\right] \left[m C(v,u)^{m-1}
R(v,u)\right] \left[ C(t,u)^{n} \right]
\nonumber\\ &&\phantom{
\times (}{}+ \left. \left[ C(t,v)^{l} \right]\left[m C(u,v)^{m-1}
R(u,v)\right] \left[ n C(t,u)^{n-1} R(t,u)\right] \right).
\label{dyn1}
\end{eqnarray}

We now exploit the ultrametric relation among fixed points of the
application $f(q,q')$ and the
constancy of $X(q)$ for values of $q$ between contiguous fixed points
that we discussed in
section~\ref{offeq}.  We can thus express $\lim_{t\to\infty}
\left<{O}(t) \right>$ as a functional of $X_\eps(q)$, namely
\begin{eqnarray}
&&\lim_{t\to\infty}E_J E_{J'} \left<{O}(t) \right> =\nonumber\\
&&\qquad{}\beta^2
\epsilon \epsilon' \frac{N}{p} \((1 - \int dq \,\frac{d X_\eps
(q)}{dq}\, \left[ q^{l+m} + q^{m+n} + q^{n+l} \right] +\int dq\,
\frac{d X_\eps (q)}{dq}\, X_\eps(q)\, q^{l+m+n} \right. \nonumber
\\ &&\qquad\phantom{\beta^2
\epsilon \epsilon' \frac{N}{p}}\left.+\int dq_1\, \frac{d X_\eps
(q_1)}{dq_1} \int_0^{q_2} dq_2 \frac{d X_\eps (q_2)}{dq_2}\,
\left[q_1^l q_2^{m+n} + q_1^m q_2^{n+l}+ q_1^n q_2^{l+m}\right]
\right). \label{corrum}
\end{eqnarray}

The derivation is given in Appendix~\ref{ultr:app}.
Since the observable $O$ is local
(in a suitably folded space), it is reasonable to expect that it
equilibrates, and thus that the long-time limit of $\la O(t)\ra$
is equal to its equilibrium expectation
value. Notice however that $O$ is \textit{not} the energy density, so that
its thermalization is not guaranteed by general principles: this is
the reason why the link between dynamic and static ultrametricity,
although very plausible, is not completely proved by our derivation.
Assuming the equilibration of $O$, we obtain
\begin{eqnarray}
P^{(3)}_\epsilon (q_{12},q_{13},q_{23})&=& \frac{1}{2}\,P_\eps
(q_{12})\, x(q_{12})\,
\delta(q_{12}-q_{13})\,\delta(q_{12}-q_{23})\nonumber\\
&&{}+\frac{1}{2}\left[ P_\eps (q_{12})P_\eps
(q_{13})\,\theta(q_{12}-q_{13}) \,\delta(q_{13}-q_{23})+{\rm
permutations}\, \right]. \label{pum}
\end{eqnarray}
This implies static ultrametricity with a weighting of the triangles as
first found in \cite{MPSTV}
from replica theory.

We shall now use again linear response to show that
dynamical ultrametricity (and
therefore, according to the above discussion, static ultrametricity)
can be related to the property of ``separability" defined in \cite{PARI}.

In order to do so, let us introduce a different measure of
overlap among states. Consider a system perturbed with
the term $ H_2({\bf S})= \sum_{x\in {\cal S}_1} J_x^{(2)} S_x S_{{\cal
T}(x)}$.
For this system we introduce a new static overlap probability
function by the usual procedure, discussed in sect.~\ref{Equilibrium}.
We introduce
two copies of the system, and define a new overlap $Q^*$ between
them by
\begin{equation}
Q^*({\bf S},{\bf S}')=\frac 1 2 \sum_{x\in {\cal S}_1} J_x^{(2)}
\(( S_x S'_{{\cal
T}(x)}+S'_x S_{{\cal
T}(x)}
\)) .
\end{equation}
The new overlap probability function $P^*(q^*)$ is defined
in complete analogy to
(\ref{pidiqu}) as the probability to find $Q^*=q^*$, at equilibrium.
Correspondingly in dynamics one can define a new correlation function
\begin{equation}
C^*(t,t')=\frac 1 2 \sum_{x\in {\cal S}_1} J_x^{(2)}
\langle  S_x(t) S_{{\cal
T}(x)}(t')+S_x(t') S_{{\cal
T}(x)}(t)\rangle,
\end{equation}
a new response function
\begin{equation}
R^*(t,t')=\frac{1}{ 2} \sum_{x\in {\cal S}_1} J_x^{(2)}
\left[  \frac {\delta\langle S_{{\cal
T}(x)}(t)\rangle }{\delta h_x(t')}+
\frac{\delta\langle
S_x(t)\rangle}{\delta h_{{\cal
T}(x)}(t')}\right],
\end{equation}
and a new FDR $X^*(q^*)$ according to the definition
(\ref{fdrat}),
with $C$ and $R$ substituted by $C^*$ and $R^*$.
$C^*(t,t')$ represents the average correlation of a spin at time $t$
with (a part of) its molecular field at time $t'$. As the correlation
induced by $H_2$ is positive, $C^*(t,t')$ should be, as $C(t,t')$, a
monotonically decreasing function of the time separation.
We can then define
for long times, a function
\[
q^*(q)=\lim_{ {t,t'\to\infty} \atop C(t,t')=q }
C^*(t,t'),\]
which is monotonic, with the exception at most
of  constant or vertical parts. These parts would represent
a completely different evolution of
$C^*(t,t')$ and $C(t,t')$, what seems to us a quite strange possibility.
This suggests that in statics pairs of states with equal $q$
also have  equal $q^*$. The algebraic expression of this property
within the replica formalism has been
discussed in \cite{PARI}.
If we admit this
property of separability,
we have
$P^*(q^*)\,dq^*=P(q)\,dq$, and via stochastic stability $X^*(q^*(q))=X(q)$.
Integrating by parts over $J_x$ as usual,  we find the following
expressions for $C^*$ and $R^*$:
\begin{eqnarray}
C^*(t,u)&=&\beta\eps\left[\int_0^u ds\; C(t,s) \, R(u,s) + R(t,s) \, C(u,s)
+\int_u^t ds \; R(t,s) \, C(s,u)\right];
\nonumber
\\
R^*(t,u)&=&\beta\eps\left[\int_u^t ds \; R(t,s) \, R(s,u)\right].
\end{eqnarray}
It is easy to check that in any correlation domain ${\cal D}_u$
where the correlation functions depend only on $h_u(t')/h_u(t)$,
the function
$C^*$ depends on the same combinations of the time variables.
If in addition we require that
$R^*(t,t')=X(C)\left({\partial C^*(t,t')}/{\partial t'}\right)$, a detailed
calculation shows that
the only possibility is that $X(q)$ is a constant in ${\cal D}_u$.

We see then that if we assume the property of separability in statics,
stochastic stability implies the existence of a unique dynamic function $X(q)$
independent of the particular
correlation and response function used in the definition.
This would corroborate the
interpretation of $T/X(q)$ as a well-defined effective temperature
for the system \cite{teff}.
This conclusion depends on the validity of
separability in the statics.  Unfortunately, we have not
been able to find a way to prove this property,
although we feel that it should be possible to advance some
arguments in favor of it, similar to the ones used above.
The connection between separability and stochastic stability is
an interesting point which needs to be clarified by further
work.

\section{Summary and comments}
\label{conclusion}
We have thus developed the linear response theory for glassy systems,
based on the equilibration of
the free energy density.  The introduction of local random perturbations
of the Hamiltonian, treated
in linear response,  naturally leads to a new equilibrium order
parameter function $\tilde P(q)$
and to a new fluctuation dissipation ratio $\tilde X(q)$, which are
obtained as the limits of the
corresponding quantities when the perturbation gets small.  From a
theoretical point of view these
two quantities possess appealing properties:
\begin{itemize}
\item They are directly related by $d \tilde X/dq=\tilde P(q)$.
Thus the link between the static equilibrium order parameter and the out of
equilibrium
dynamics is clear, and $\tilde P(q)$ can
be measured by performing an off equilibrium dynamical measurement.
\item If the original Hamiltonian $H_0$ has some quenched-in disorder,
the static order parameter function $\tilde P(q)$ is self-averaging with
respect to this disorder.
\item The order parameter function $\tilde P(q)$ is defined by the
values of extensive, thermodynamic quantities.
\end{itemize}
Therefore linear response allows to define a single, self
averaging, order parameter
function which is able to describe both the equilibrium properties and
the off-equilibrium
dynamics.

If stochastic stability holds, i.e., if the average correlation
functions are continuous with
respect to small perturbations of the system, $\tilde{P}(q)$ is equal to
the overlap probability
function $P(q)$ of the unperturbed system,
up to some simple effects
of global symmetries which we have discussed.  We have thus found the root
of the formal identities
between static and off-equilibrium dynamical quantities, often
empirically found in the analysis of
specific models.  In general, if the system has a symmetry which is
lifted by the perturbation, like
a global Ising symmetry for a spin glass in zero field, then the new
order parameter $\tilde P(q)$
is equal to the usual overlap probability distribution in the presence
of an infinitesimal breaking.

The discussion has been extended to more complex characterization of the
statistics of the equilibrium states. Stochastic stability allows to infer
properties of the dynamics from properties of the statics and vice-versa.
In this way we have
argued, modulo some technical
assumptions,
 that if {\it static} separability holds, i.e.
if one can not
define different notions of overlap yielding different predictions on
the relation among the states, then {\it dynamically}
the fluctuation dissipation ratio is  a uniquely defined function
independent on the particular correlation and response functions used in the
definition. This
suggests that ultrametricity should hold in all stochastically stable
systems, both in the out of equilibrium dynamics
and in the equilibrium properties.

To conclude, let us stress the important experimental implications
of this work.  Whenever the values of the energy density can be
considered close to their equilibrium, the order parameter function
$P(q)$ can be obtained from the
determination of the function $X(q)$ in an aging experiment.  This
requires separate measurements
of the response and the fluctuation, which are in principle both
accessible to experimental
determination.  Although generically valid for infinite time in short
range systems, depending on
the specific situation, the energy density equilibration can be fulfilled
or fail on laboratory time
scales.  In structural glasses for example, the relaxation time for
energy density is so large that
one is never probing the time region to which our analysis apply.  In
spin glasses conversely it is
generally believed that energy densities are asymptotically close to
equilibrium.  An experimental
determination of $X(q)$ in spin glasses would be therefore of the utmost
interest, being equivalent
to the determination of $P(q)$.

\section*{Acknowledgments}
We thank F. Ricci-Tersenghi and
M.A. Virasoro for stimulating discussions.
The work of MM has been supported in part by the
National Science Foundation under grant No.~PHY94-07194.

\appendix

\section{}\label{dyn:app}
We now derive the equalities (\ref{SLR}) and (\ref{DLR}). We start
from the identity, valid for any set of Gaussian independent
variables $J_{i_1,\ldots,i_p}$ and for any function $f(J)$:
\begin{equation}
 E_J [ J_{i_1,\ldots,i_p}f(J)]=E_J[ J_{i_1,\ldots,i_p}^2]\, E_J
 \left[\frac{\partial f(J)}{\partial J_{i_1,\ldots,i_p}} \right].
\end{equation}
Let us consider the canonical average of $H_p^{\rm LR}$,
\begin{eqnarray}
E_J\langle H_p^{\rm LR}\rangle &=&
E_J\left[\sum_{i_1<\cdots<i_p}J_{i_1,\ldots,i_p}\,\left<S_{i_1}\cdots
 S_{i_p}\right>\right]\nonumber\\
 &=& \frac{p!}{N^{p-1}}
\sum_{i_1<\cdots<i_p}E_J\left[ \frac{\partial}{\partial
J_{i_1,\ldots,i_p}} \left\langle S_{i_1}\cdots S_{i_p}\right
\rangle \right]\nonumber\\ &=& -\frac{\beta\epsilon p!}{N^{p-1}}
 \sum_{i_1<\cdots<i_p}E_J \left[1- \left\langle S_{i_1}\cdots
 S_{i_p}\right\rangle^2\right] \nonumber\\ &=& - N \beta\epsilon
\left(1-\int dq \; P(q)\; q^p\right).
\end{eqnarray}
We obtain analogously in Langevin dynamics:
\begin{equation}
E_J\langle H_p^{\rm LR}(t)\rangle = \frac{p!}{N^{p-1}}
\sum_{i_1<\cdots<i_p}E_J  \left[\frac{\partial}{\partial
J_{i_1,\ldots,i_p}} \left\langle S_{i_1}(t)\cdots S_{i_p}(t)
\right\rangle\right].
\end{equation}
By considering the expression (\ref{MSR:action}) of the
Martin-Siggia-Rose action we obtain
\begin{eqnarray}
E_J\langle H_p^{\rm LR}(t)\rangle
&=&\frac{\eps\,p!}{N^{p-1}}\int_0^t dt'\;\sum_{i_1<\cdots<i_p}
p\,\left\langle S_{i_1}(t)\cdots S_{i_p}(t)
i\hat{S}_{i_1}(t'){S}_{i_2}(t')\cdots
S_{i_p}(t')\right\rangle\nonumber\\& =& N\eps \int_0^t dt'\; p
\,C^{p-1}(t,t') R(t,t') .\label{a3}
\end{eqnarray}
As stated in the text, equation (\ref{a3}) holds in a much more
general dynamical context. Consider a general  relaxational
dynamics in which spins are updated according to the local field,
e.g.,
\begin{equation}
S_x(t+\delta t)= f(h_x(t),\eta_x(t)),
\end{equation}
where $h_x(t)$ is the molecular field at time $t$ (which depends
on the other spins and on the quenched variables) and $\eta_x(t)$
represents the thermal noise.  $S_x(t)$ depends on
$J_{i_1,\ldots,i_p}$ only through its implicit dependence on
$h_x(t')$ at all previous times. Therefore,
\begin{eqnarray}
E_J \langle J_{i_1,\ldots,i_p} S_{i_1}(t)\cdots S_{i_p}(t) \rangle
&=&E_J(J^2)\, E_J  \frac{\partial}{\partial J_{i_1,\ldots,i_p}}
\left\langle S_{i_1}(t)\cdots S_{i_p}(t)\right\rangle\nonumber\\ &
= & E_J(J^2)\,E_J\int_0^t dt' \; \sum_j \left\langle
\frac{\partial h_j(t')} {\partial J_{i_1,\ldots,i_p}}
\frac{\partial}{\partial h_j(t')} S_{i_1}(t)\cdots S_{i_p}(t)
\right\rangle.
\end{eqnarray}
Now it is easy to see that
\begin{equation}
\frac{\partial h_j(t')} {\partial J_{i_1,\ldots,i_p}} =
\sum_{\ell=1}^p \delta_{j,i_\ell} S_{i_1}(t')\cdots
\{S_{i_\ell}(t')\}\cdots S_{i_p}(t'),
\end{equation}
where $S_1\cdots \{ S_\ell \} \cdots S_p$ means that $S_\ell$ does
\textit{not} appear in the product. We thus find the relation
(\ref{a3}).

\section{}\label{ultr:app}
In this appendix we discuss the technical derivation of formulae
(\ref{stat1}) and (\ref{dyn1}),
together with the the prove of (\ref{corrum}) if dynamical
ultrametricity holds.  Although
considerably more involved the derivations parallel the ones for the
$P(q)$ and $X(q)$ without
adding new physical insight.  In order to simplify the exposition we
introduce a multi-index
notation
\begin{eqnarray}
i(x)&=&\{x, {\cal T}_1^{(p)}(x),
\ldots, {\cal T}_{l-1}^{(p)}(x)\};
\nonumber
\\
j(x)&=&\{{\cal T}_l^{(p)}(x),
\ldots, {\cal T}_{l+m-1}^{(p)}(x)\};
\nonumber
\\
k(x)&=&\{{\cal T}_{l+m}^{(p)}(x),
\ldots, {\cal T}_{p-1}^{(p)}(x)\};
\end{eqnarray}
and
\begin{eqnarray}
{\bf S}_{i(x)}&=&S_x S_{{\cal T}_1^{(p)}(x)}
\ldots, S_{{\cal T}_{l-1}^{(p)}(x)};
\nonumber
\\
{\bf S}_{j(x)}&=&S_{{\cal T}_l^{(p)}(x)}
\ldots, S_{{\cal T}_{m+l-1}^{(p)}(x)};
\nonumber
\\
{\bf S}_{k(x)}&=&S_{{\cal T}_{m+l}^{(p)}(x)}
\ldots, S_{{\cal T}_{p-1}^{(p)}(x)}.
\end{eqnarray}
In these notations one has
\begin{eqnarray}
H_p^{(1)}&=&\sum_x J_x {\bf S}_{i(x)}{\bf S}_{j(x)},\\
H_p^{(2)}&=&\sum_x J_x {\bf S}_{j(x)}{\bf S}_{k(x)},\\
O&=&\sum_x J_x J'_x {\bf S}_{i(x)}{\bf S}_{k(x)}.
\end{eqnarray}
If we compute the average of $O$ in statics we obtain
\bea
E_J E_{J'}\langle O\rangle &=&\sum_{x\in {\cal S}_1}
E_J E_{J'}\frac{\partial}{\partial J_{x}}\frac{\partial}{\partial
J'_{x}}
\langle
{\bf S}_{i(x)} {\bf S}_{k(x)}
\rangle \\
&=&
\beta^2 \eps\eps' E_J E_{J'}\left(
1-
\langle
{\bf S}_{i(x)} {\bf S}_{k(x)}
\rangle^2
-\langle
{\bf S}_{j(x)} {\bf S}_{k(x)}
\rangle^2
-
\langle
{\bf S}_{i(x)} {\bf S}_{j(x)}
\rangle^2\right.\nonumber\\
&&\phantom{\beta^2 \eps\eps' E_J E_{J'}}\left.{}+2
\langle
{\bf S}_{i(x)} {\bf S}_{k(x)}
\rangle
\langle
{\bf S}_{j(x)} {\bf S}_{k(x)}
\rangle
\langle
{\bf S}_{i(x)} {\bf S}_{j(x)}
\rangle \))
\eea
which, inserting the definition of $P(q)$ and $P^{(3)}(q,q',q'')$,
reduces to (\ref{stat1}).

In dynamics as well the derivation can be simplified by the use of a
smart notation.
The perturbations $H_1$ and $H_2$ induce
the following terms respectively in the dynamical action:
\begin{equation}
\int_0^t d u \sum_{x\in {\cal S}_1}J_x \sum_{r=0}^{l+m-1}
\left[ S_x(u)\cdots \{{S}_{\cT_r(x)}(u)\} \cdots {S}_{\cT_{l+m-1}(x)}(u)
\right]
i\hat{S}_{{\cal T}_r^{(p)}(x)}(u),
\label{uno}
\end{equation}
and
\begin{equation}
\int_0^t d u \sum_{x\in {\cal S}_1} J_x'\sum_{r=l}^{p-1}
\left[ S_{{\cT_{l}(x)}}(u)\cdots \{{S}_{\cT_r (u)}\}\cdots
{S}_{\cT_{p-1}}(u)\right]
i\hat{S}_{\cT_r{(u)}}.
\label{due}
\end{equation}
Let us now define
%\begin{equation}
\begin{eqnarray}
i{ \bf {\hat{S}} }_{i(x)}
&=&\sum_{r=0}^{l-1}
\left[ S_{x}(u)
\cdots
\{
{S}_{\cT_r}(u)
\}
\cdots
{S}_{\cT_{l-1}}(u)\right]
i\hat{S}_{\cT_r(u)};
%\end{equation}
%\begin{equation}
\nonumber
\\
i{\bf {\hat{S}} }_{j(x)}
&=&\sum_{r=l}^{l+m-1}
\left[ S_{\cT_{l}(x)}(u)
\cdots
\{
{S}_{\cT_r(x)}(u)
\}
\cdots
{S}_{\cT_{l+m-1}(x)}(u)\right]
i\hat{S}_{\cT_r (x)}(u);
%\end{equation}
%\begin{equation}
\nonumber
\\
i{\bf {\hat{S}}}_{k(x)}
&=&\sum_{r=l+m}^{p-1}
\left[S_{\cT_{l+m}(x)}(u)
\cdots
\{
{S}_{\cT_r(x)}(u)
\}\cdots
{S}_{\cT_{p-1}(x)}(u)\right]
i\hat{S}_{\cT_r(x)}(u).
%\end{equation}
\end{eqnarray}
With this notation the terms (\ref{uno}) and (\ref{due})
are respectively written  as
\begin{eqnarray}
&& \int_0^t du \; \sum_{x\in {\cal S}_1}J_x \left[ i{\bf {\hat
S}}_{i(x)}(u){\bf S}_{j(x)}(u) +i{\bf {\hat S}}_{j(x)}(u){\bf
S}_{i(x)}(u)\right];
\nonumber
\\
&& \int_0^t du \; \sum_{x\in {\cal S}_1}J'_x \left[ i{\bf {\hat
S}}_{j(x)}(u){\bf S}_{k(x)}(u) +i{\bf {\hat S}}_{k(x)}(u){\bf
S}_{j(x)}(u)\right].
\end{eqnarray}
Evaluating $\langle O(t)\rangle $ by integration by parts
we obtain
\begin{eqnarray}
\langle O(t)\rangle =&&\eps\eps' \sum_{x\in {\cal S}_1}
\int_0^t du \int_0^t dv \left\langle
{\bf S}_{i(x)}(t){\bf S}_{k(x)}(t)\left[
i{\bf {\hat
S}}_{i(x)}(u){\bf S}_{j(x)}(u) +i{\bf {\hat S}}_{j(x)}(u){\bf S}_{i(x)}(u)
\right] \right.
\nonumber
\\
\times &&
\left.
\left[
i{\bf {\hat
S}}_{j(x)}(v){\bf S}_{k(x)}(v) +i{\bf {\hat S}}_{k(x)}(v){\bf S}_{j(x)}(v)
\right] \right\rangle.
\end{eqnarray}
Using the clustering property for far-away sites for small $\eps$,
$\eps'$,
we get the relation
\begin{equation}
\langle {\bf S}_{i(x)}(t)i{\bf {\hat
S}}_{i(x)}(u) \rangle =
l C^{l-1}(t,u) R(t,u),
\end{equation}
and analogous ones for the averages involving the multi-indices
$j(x)$ and $k(x)$, which collected together give eq.~(\ref{dyn1}).

Let us now sketch the derivation of
eq.~(\ref{corrum}) from dynamical ultrametricity.
Let us simplify the notations in eq.~(\ref{dyn1}) by writing, for
$r=l,m,n$,
\begin{eqnarray}
C_r(t,u)&=&C^r(t,u); \nonumber\\
R_r(t,u)&=&r C^{r-1}  (t,u) R  (t,u).
\end{eqnarray}
In this way eq.~(\ref{dyn1}) reads:
 \begin{eqnarray}
\frac{E_J E_{J'} \left<{O}(t) \right>}{\beta^2 \epsilon \epsilon' N}
&=&
\frac{1}{p} \int_0^t dv \int_0^{t} du\;
\left[
R_l(t,v)C_m(v,u)R_n(t,u)\right.\nonumber\\
&&\qquad \left.{}+ C_l(t,v)R_m(v,u)R_n(t,u)
+ C_l(t,v)R_m(u,v)R_n(t,u)
 \right].
\label{dyn1app}
\end{eqnarray}
For long times we can write $R_r(t,u)\approx X_\epsilon (C(t,u))\partial_u
C_r(t,u)$, and the l.h.s.\ of eq.~(\ref{dyn1app})
becomes:
\begin{eqnarray}
\frac{1}{p} \int_0^t du \int_0^{u} dv
& &\,\left[
X_\epsilon (C(t,v))\partial_v C_l(t,v)C_m(v,u) X_\epsilon (C(t,u))\partial_u 
C_n(t,u)
\right.
\nonumber\\
& &
\quad\left.
{}+ C_l(t,v)X_\epsilon (C(u,v))\partial_u C_m(u,v)X_\epsilon (C(t,u))\partial_u 
R_n(t,u)
 \right] + \left[n \leftrightarrow l \right].
\label{dyn2app}
\end{eqnarray}

  The fact that $X_\epsilon (C)$ is constant within each crossover
domain allows as to manipulate this equation as if the
ultrametric relation among the correlation function were to hold for all
time involved. This can be realized as the constancy of $X_\epsilon $ implies
that in each domain  one has to integrate a total derivative and
the integrals can be performed explicitly.

We now introduce the shorthand notations
\begin{eqnarray}
(\partial C(t,u)/\partial u)du & \longmapsto & dC(t,u);\\
n C^{n-1}(t,u) dC(t,u) &\longmapsto & dC_n(t,u),
\end{eqnarray}
and similarly for the other differential.
Using this notation and the constancy of $X_\epsilon$,
we find that the first term of
eq.~(\ref{dyn2app}) can be written as
\begin{equation}
\int_0^1 dC_n(t,u)\, X_\epsilon (C(t,u))\,\int_0^{C(t,u)}
d C_l(t,v)\; C_m(t,v)\, X_\epsilon (C(t,v)),
\end{equation}
while the second one is given by
\begin{equation}
\int_0^1 dC_m(t,v)\, X_\epsilon (C(t,v))\,\int_0^{C(t,v)}
d C_n(t,u)\; C_l(u,v)\, X_\epsilon (C(u,v))
+[m \leftrightarrow n].
\end{equation}
The extrema of integration refer to the variable $C$.
Integrating by parts and collecting all terms we find
eq.~(\ref{dyn1}).

\end{document}